\documentstyle[12pt,aasms4]{article}

\def\la{\mbox{\raisebox{-0.1ex}{$\scriptscriptstyle \stackrel{<}{\sim}$\,}}}

\newcommand{\alphaone}{\mbox{$\alpha_{1}$}}
\newcommand{\alphatwo}{\mbox{$\alpha_{2}$}}
\newcommand{\betaone}{\mbox{$\beta_{r1}$}}
\newcommand{\betatwo}{\mbox{$\beta_{r2}$}}
\newcommand{\deltaS}{\mbox{$\Delta s $}}
\newcommand{\deltaSvec}{\mbox{$\bf {\Delta s }$}}
\newcommand{\deltaSmin}{\mbox{${\Delta s}_{min}$}}
\newcommand{\deltaph}{\mbox{$\Delta \phi $}}
\newcommand{\degsp}{$^{\circ\/\ }$}
\newcommand{\fc}{\mbox{$f_{0}$}}
\newcommand{\Lp}{\mbox{$L_{p}$}}
\newcommand{\Ls}{\mbox{$L_{s}$}}
\newcommand{\mupm}{\mbox{$\mu_{pm}$}}
\newcommand{\nut}{\mbox{$\nu_{t}$}}
\newcommand{\nuf}{\mbox{$\nu_{f}$}}
\newcommand{\thetaone}{\mbox{$\theta_{r1}$}}
\newcommand{\thetatwo}{\mbox{$\theta_{r2}$}}
\newcommand{\thetaonevec}{\mbox{$\bf {\theta_{r1}}$}}
\newcommand{\thetatwovec}{\mbox{$\bf {\theta_{r2}}$}}

\newcommand{\thetadiff}{\mbox{$\theta_{r12}$}}
\newcommand{\rvec}{\mbox{\bf r}}
\newcommand{\Viss}{\mbox{$V_{iss}$}}

\begin{document}


\title{\Large\bf Multiple Imaging of PSR B1133+16 by the ISM } 

\vspace{5.0cm}

\author{\normalsize\bf Yashwant Gupta\footnote{send preprint requests to 
$ygupta@ncra.tifr.res.in$}, N. D. Ramesh Bhat, and A. Pramesh Rao } 


\begin{center}
{\normalsize National Centre for Radio Astrophysics, Tata Institute of Fundamental Research, \\
Post Bag 3, Ganeshkhind, Pune - 411 007, India}
\end{center}

\vspace{5.0cm}

\begin{center}
{\normalsize\bf Accepted for publication in The Astrophysical Journal \\
(Vol 520 \#1, July 20, 1999)}
\end{center}


\begin{abstract}

Refraction of pulsar radiation by electron density irregularities in the 
interstellar medium sometimes produces multiple imaging of pulsars 
which can lead to periodic oscillations of intensity in pulsar dynamic 
spectra records.  Such events can be used as tools to resolve the emission 
regions in pulsar magnetospheres.  
Here we describe results from the recent observation of a double imaging 
event for PSR B1133+16, which place fairly tight constraints on the location
of the emission regions. 
Our analysis constrains the location of the scattering object to the shell 
of the Local Bubble. 
The phase of the oscillations shows significant variations across the
pulse.   The minimum value for the transverse separation of the emitting 
regions at the two edges of the pulse is inferred to be $3 \times 10^5 $ m.  
This translates to a minimum emission altitude of $ 2.6 \times 10^6 $ m.
The non-monotonic variations of the fringe phase with pulse longitude are
interpreted as variations of the altitude of the emission regions for the
orthogonal polarization modes of this pulsar.  This is in agreement with
theories where propagation effects, such as refraction, are responsible
for the orthogonal modes. 

\end{abstract}



\section{Introduction}

Random electron density fluctuations in the ionized component of the 
interstellar medium (ISM) produce interstellar scintillation (ISS) of the 
radio signals received from pulsars.
The small-scale irregularities ($ \sim 10^{7} $ m) in the ISM produce a 
scatter broadened image of the pulsar, characterized by the diffractive 
scattering angle, $\theta_d$. 
The large-scale irregularities ($\sim 10^{11} $ m) steer the diffractive 
scattering cone by a mean direction of arrival, $\theta_r$, which is
expected to vary on refractive scintillation time scales.  In dynamic 
spectra observations of pulsars, diffractive scintillation shows up as the 
random modulation of intensity across the time-frequency plane. 
The sloping nature of these intensity features at a single epoch is due 
to the refractive steering (see Rickett 1990 for a review of diffractive
and refractive scintillations).

Sometimes, strong refraction effects in the ISM produce two or more 
scatter broadened images of the pulsar with well separated directions of 
arrival, $\thetaone$ and $\thetatwo$. Such multiple imaging events produce 
additional features in pulsar dynamic spectra, such as multiple 
drift slopes and periodic intensity modulations or fringes, over and above 
the random modulations due to diffractive scintillation (e.g. Cordes \& 
Wolszczan 1986; Gupta, Rickett \& Lyne 1994).
Study of multiple imaging events allows a better understanding of the 
distribution of electron density irregularities in the ISM, particularly at 
the large scales (e.g. Rickett 1996). An example of the application of
this technique is the inference of enhanced scattering from a cloud of warm
ionized medium from the analysis of a multiple imaging event detected for 
PSR B0834+06 (Rickett, Lyne \& Gupta 1997).

Multiple imaging events also provide an exciting technique for studying pulsar 
emission geometry, as they allow the possibility of resolving the pulsar 
magnetosphere.  This happens because the phase of the fringes is sensitive 
to the transverse location of the source of radiation.  Observed variations
of this phase as a function of pulsar longitude can be used to infer the
transverse separation of the emitting regions across the polar cap.
This method was first used by Wolszczan \& Cordes (1987) to resolve the
emission region for PSR B1237+25 and later on by Wolszczan, Bartlett \& 
Cordes (1988) for PSR B1133+16 and more recently by Kuz'min (1992) for 
PSR B1919+21.  From measurements of the phase change of the fringes from 
the leading to the trailing edge of the pulse, these authors report values 
ranging from $ 5 \times 10^6 $ m to $3 \times 10^7 $ m for the transverse 
separation of the emission regions in the magnetosphere.  For a dipole 
field geometry with constant emission altitude across the polar cap, these 
values translate to emission altitudes which are of the order of the light 
cylinder distance ($ 10^7 - 10^8 $ m).  
These estimates of emission altitudes are typically one to two orders of 
magnitude more than those estimated from other techniques such as 
period-pulse width relationships, multi-frequency timing (see Cordes 1992 
for a summary), which infer emission altitudes in the range $10^5 - 10^6$ m.  
In order to obtain an improved understanding of the location of emission
regions in pulsar magnetospheres, it is important to resolve these 
differences.  More accurate scintillation measurements that place tighter
constraints on the size and location of emission regions can be very
useful in this context.

In this paper, we present the results from new observations of multiple imaging
of the pulsar PSR B1133+16.  The observations and analysis of the data are described
in section 2.  In section 3, we describe the model used to explain the multiple
imaging episode and infer the size of the emission region for the pulsar from this.
Section 4 discusses the implications of our results for the distribution of
emission regions in pulsar magnetospheres.  Our conclusions are summarized in
section 5.


\section{Observations and Data Analysis}

As part of a regular program for monitoring slow variations of diffractive 
scintillation properties of pulsars due to refractive scintillation, PSR B1133+16 
was observed at 27 epochs from March 1994 to May 1994, using the Ooty Radio 
Telescope (Bhat, Rao \& Gupta 1999a). The observations were 
carried out at a frequency of 327 MHz with a 9 MHz bandwidth spanned by 64 
spectral channels. The dynamic spectra were generated by on-line averaging of 
10 successive pulses, leading to an effective time resolution $\approx$ 12 sec.  
The data were gated around the on-pulse phase of the pulsar signal with a 
sampling interval of 6 msec, thereby providing 10 phase resolved dynamic 
spectra with a resolution of 1.8 degrees of longitude.  The details of the 
data analysis procedure are described in Bhat, Rao \& Gupta (1999a).

Dynamic spectra at three epochs $-$ 28 April 1994, 30 April 1994 and 
2 May 1994 $-$ in the observing session are shown in Figure 1.  The dynamic
spectra shown here are the sum from all the 10 longitude bins.  There are
a maximum of 512 time samples in these plots, and of the 64 frequency channels, 
the display is restricted to channels 8 to 58.  The intensity is represented 
on a linear grey scale with darker regions corresponding to higher intensity.  
The mean pulsar flux density for each epoch is indicated in the panels.  The 
signal to noise ratio for these data is very good.  For example, the mean signal 
to noise ratio for a data point in the dynamic spectrum for 30 April 1994 is 
estimated to be $\approx 23$.  The signal to noise ratio for the bright intensity 
scintles will clearly be 3 to 4 times more than this.  When the dynamic spectra 
data is split into individual longitude bins, this signal to noise ratio is 
reduced, to a first order, by a factor of $\sqrt{10}$, giving a mean signal to 
noise ratio for single bin data of 7.3.

The broad-band vertical white stripes in 
the display are due to residual intrinsic pulse-to-pulse intensity modulations 
of the pulsar that are not smoothed out in the 10 period averages.  The random 
intensity fluctuations produced by diffractive ISS are clearly seen at all the 
three epochs.  Though located randomly, these intensity modulations (commonly 
referred to as scintles) have fairly well defined slopes in the frequency-time 
plane, which is due to the effect of large-scale, refractive phase gradients 
produced by the scattering medium (e.g. Hewish 1980).  At all the three epochs, 
the predominant slope is positive (i.e., scintles are aligned along increasing 
frequency and time).  However, there is indication, in the first two epochs, of 
the presence of some scintles with an opposite slope.  Though single 
drift slopes are the norm, such occurrences of dual slopes in pulsar dynamic 
spectra are not uncommon (e.g. Hewish 1980). They are typical in cases where 
there are two well separated directions of arrival for the refracted wavefronts. 

A closer examination of the data from the second epoch (30 April 1994) reveals 
that in addition to the random intensity fluctuations due to diffractive ISS, 
periodic intensity modulations that are finer than the typical widths of the 
scintles are also present.
Such periodic intensity modulations, commonly referred to as fringes, are 
typical indicators of the multiple imaging phenomenon.  The data from the 
first and third epochs show no sign of such fringes, indicating that the 
phenomenon lasted for a fairly short duration.

A secondary spectral analysis of the dynamic spectra data was carried out by 
performing a 512 x 64 point two-dimensional Fourier Transform. 
The data were zero padded to take care of the frequency channels rejected at 
the edges of the band.  A periodicity in the primary dynamic spectra should
show up as a significant peak in the secondary spectra.  For the data from
the second epoch, a clear peak was found corresponding to a periodicity of 
\begin {equation}
\nut ~=~ 5.9 \times 10^{-3}~~{\rm cycles/sec ~~~and~~~}  \nuf ~=~ 1.6 ~ {\rm cycles/MHz}
\end {equation}
in the inverse time and inverse frequency domains, respectively.  However, 
the peak was not confined to a single bin in the secondary spectrum, 
but was spread over about 5 bins along the inverse time axis and 4 bins along 
the inverse frequency axis.

To further elucidate the nature of the periodicities in the dynamic spectra, 
the data were put through a band pass filter centred around the frequency 
of the periodic signal.  The results are shown in the top panel of Figure 2 where 
the detailed nature of the periodicities is clearly brought out.  The amplitude of 
the fringes is modulated by the amplitude of the diffractive scintles and the
phase shows discrete jumps at the scintle boundaries.  Both these effects
reinforce the interstellar scintillation origin of the fringe pattern.
These two effects are also responsible for the observed broadening of the 
spectral peak in the Fourier Transform domain.
The data were also put through a low pass filter that rejected the frequency
corresponding to the periodic signal.  The result is shown in the middle panel
of Figure 2.  Here, the intensity grey scale is a logarithm scale which has been
used to enhance the low intensity features at the expense of the high intensity
features.
It is clear from this figure that the dynamic spectra data show dual drift 
slopes, which is naturally expected when periodic fringe patterns are 
produced.  The primary slope is the easily discernible positive slope (intensity
scintles drifting along increasing frequency with increasing time). The secondary 
slope is a negative slope that is less prominent than the positive slope but can 
be seen in the form of "V" shaped intensity structures as well as several individual 
scintles showing a negative slope.
This is further exemplified by the autocorrelation function (ACF) plots 
shown in the bottom panel of Figure 2.  The first of these is for the normal,
low pass filtered data while the second is for the log-scaled, low pass filtered
data.  The ACF for the normal data is dominated by the primary drift slope whereas
the ACF for the log-scaled data clearly shows the presence of dual drift slopes.
The solid and dashed lines on these plots mark the two estimated drift slopes whose
values are found to be
\begin {equation}
\frac {dt} {df} \biggl |_{1} ~~=~~  142 ~\pm~ 25 ~{\rm sec/MHz}  ~~~{\rm and}~~~ 
\frac {dt} {df} \biggl |_{2} ~~=~~ -219 ~\pm~ 50 ~{\rm sec/MHz}  ~~.
\end {equation}
The error bars are 1-$\sigma$ values, as estimated in Bhat, Rao \& Gupta (1999a).

Our time resolution of 6 msec provided 10 time samples across the pulse width.
Analysis showed the fringes to be present in each of these 10 phase-resolved 
dynamic spectra, with the average amplitude of the periodicity following the 
mean pulse shape of the pulsar (Fig. 3, upper panel).  
For the dynamic spectra at each longitude bin, the band pass filtering was
done to obtain the periodic signal.  The filtered data from one reference bin 
was then cross-correlated with similar data from each bin to obtain 10 temporal
cross-correlation functions.  In the computation of these cross-correlation 
functions, the maximum lags in time and frequency domains were restricted to 
64 and 32 respectively to reduce the effect of phase jumps in the dynamic spectra,
such as those seen at scintle boundaries. 
Fourier Transforms were taken for each of these
cross-correlation functions and the phase of the signal in the frequency bin 
corresponding to the periodicity in time was extracted to give the fringe 
phase versus longitude bin curve shown in the lower panel of Figure 3.  The errors 
on these phase estimates were obtained by calculating the inverse tangent of the 
ratio of the root mean square noise in the off signal bins to the amplitude 
of the signal bin of the Fourier Transform.
The total variation of the phase of the fringes across the pulse ($\deltaph$) 
was found to be $\approx$ 15\degsp (Fig. 3, lower panel).  The phase variation was, 
however, not monotonic, showing two local extrema within the pulse which
are located at bin numbers 3 and 5.  To check the robustness of our phase curve 
and the associated error bars, we repeated the above analysis for subsets obtained 
by splitting the data into two sections of 256 samples by 64 channels each.  The 
phase curves obtained from these two subsets were found to match quite well with 
the result in Figure 3, except for the first and last longitude bins where the 
deviations were somewhat larger, but still within the estimated error bars.


\section{Modelling and Interpretation of the Observations}

Our observations and data analysis indicate that the observed fringes are 
consistent with multiple imaging due to large-scale irregularities in the ISM.  
We model the observed refractive effects as being due to scattering from a 
discrete structure, such as a plasma lens, located at a distance $\Ls$ from 
the observer along the line of sight to the pulsar (Fig. 4(a)).  Here the pulsar 
is located at P, at a distance $\Lp$ from the observer who is located at O.  
The scattering structure produces bending of radiation from the pulsar 
traveling along the directions ${\rm P - S_1 }$ and ${\rm P - S_2 }$ by angles $\betaone$ 
and $\betatwo$, respectively.  
The observer sees two dominant images of the pulsar, towards ${\rm O - S_1 }$ and 
${\rm O - S_2 }$. 
The refracted wavefronts, with angles of arrival
$\thetaone$ and $\thetatwo$, produce an interference pattern that results in
periodic intensity modulations in the space and frequency domains at O.
In addition, due to diffractive ISS, each of these two images is broadened to 
an angular spectrum of half-power width $\theta_{d}$, the diffractive scattering 
width, which produces the diffractive intensity modulations at O.  
Due to relative motion between the observer and the scintillation pattern,
quantified by the scintillation velocity $(\Viss)$, the spatial modulations
of intensity are mapped to temporal modulations.
The observation that the periodicities are much finer than the typical diffractive 
decorrelation widths in time and frequency requires that 
$ | \thetaone - \thetatwo | > \theta_{d} $.
This is equivalent to the presence of two scatter broadened, but well separated
images of the pulsar, as seen by the observer at O.
The analytical treatment of such a scenario is very similar to the description given in 
Cordes \& Wolszczan (1986) , Cordes \& Wolszczan (1988) and Rickett et al. (1997).  
However, since our final equations are somewhat different from the results
in these papers, we go through the model in some detail.

The electric field phasors at the observing plane, due to radiation from the two 
directions, can be written as 
\begin {equation}
e_{1}(\rvec,f) \ = \ a_{1}(\rvec,f)~exp[i\Phi_{1}(\rvec,f)] ~~~{\rm and}~~~
e_{2}(\rvec,f) \ = \ a_{2}(\rvec,f)~exp[i\Phi_{2}(\rvec,f)] ~~;
\end {equation}
where $a_{1}$, $a_{2}$, $\Phi_{1}$ and $\Phi_{2}$ vary randomly with $\rvec$
and $f$ due to diffractive scintillation, with typical correlation scales in
space and frequency.
The total intensity at the observing plane can then be written as
\begin {equation}
I(\rvec,f) ~~=~~ a^{2}_{1}(\rvec,f) ~+~ a^{2}_{2}(\rvec,f) ~+~ 
2~a_{1}(\rvec,f)~a_{2}(\rvec,f)~cos[\Phi_{1}(\rvec,f) - \Phi_{2}(\rvec,f)] ~~.
\end {equation}
Here the first two terms represent the diffractive scintillation patterns that would
have been produced by each of the two images alone and the third term produces the
periodic intensity modulations, due to systematic variations of $\Phi_{1} - \Phi_{2}$
as a function of $\rvec$ and $f$.
The total phase change $\Phi_{1}$ can be written as 
\begin {equation}
\Phi_{1}(\rvec,f) \ = \ \frac {k} {2} \ {(\betaone - \thetaone)}^{2} \ (\Lp-\Ls) ~+~ 
\frac {k} {2} \ \theta^{2}_{r1} \ \Ls ~+~ k \ \rvec.{\thetaonevec} ~+~ \phi_{d1}  ~~.
\end {equation}
The sum of the first two terms represents the geometric path delay due to the path 
${\rm P - S_{1} - O}$ taken by the first wavefront (with respect to the straight line 
path ${\rm P - O }$).  The third term represents the wavefront delay at the observing plane
due to the direction of arrival $\thetaone$.  The last term represents the random 
diffractive phase modulation.  A similar expression holds for $\Phi_{2}$.  Using the
relationship $\betaone \ = \ \Lp \ \thetaone/(\Lp-\Ls)$, the phase difference 
$\Phi_{1} - \Phi_{2}$ can be expressed as
\begin {equation}
\Phi_{1} - \Phi_{2} ~~=~~ \frac {1} {2} ~ \frac {k~\Ls~\Lp} {\Lp-\Ls} ~ 
(\theta^{2}_{r1} - \theta^{2}_{r2}) ~+~ k \ \rvec.(\thetaonevec - \thetatwovec) 
~+~ (\phi_{d1} - \phi_{d2})  ~~.
\end {equation}
In order to obtain the expressions for the periodicities of the fringes in frequency 
and time, we express this phase difference as  
\begin {equation}
\Phi_{1} - \Phi_{2} ~=~ 2 \ \pi \ \nuf \ df ~+~ 2 \ \pi \ \nut  \ dt ~+~ 
(\phi_{d1} - \phi_{d2}) ~+~ (\Phi_{1o} - \Phi_{2o}) ~~.
\end {equation}
Here $\Phi_{1o}$ and $\Phi_{2o}$ are the values of $\Phi_{1}$ and $\Phi_{2}$ evaluated
at the central frequency and time of observation.
Comparing with equation (6), and mapping spatial variations to temporal variations using
the scintillation pattern velocity $(\Viss)$, we have that $\nut$ and $\nuf$ are 
given by
\begin {equation}
\nut ~=~ \frac {\fc} {c} ~ \bf {V}_{iss}.(\thetaonevec - \thetatwovec) ~~,
\end {equation}
\begin {equation}
\nuf ~=~ \frac {3} {2c} ~ \frac {\Ls~\Lp} {\Lp-\Ls} ~ (\theta^{2}_{r1} - \theta^{2}_{r2}) ~~.
\end {equation}
Using the relationship between scintillation velocity and proper motion velocity
for a thin screen case (e.g. eq. [5] of Gupta 1995), the expression for $\nut$ can be 
rewritten as 
\begin {equation}
\nut ~=~ \frac {\fc} {c} ~ \frac {\Ls~\Lp} {\Lp-\Ls} \ \mupm \ 
[\thetaone \ cos(\alphaone) - \thetatwo \ cos(\alphatwo)]  ~~.
\end {equation}
Here $\mupm$ is the amplitude of the pulsar's proper motion.  The angles
$\alphaone$ , $\alphatwo$ are the angles between $\mupm$ and $\thetaonevec$ ,
$\thetatwovec$, as projected on the plane of sky.  These are illustrated in
Figure 4(b).  Also shown there is the vector $\deltaSvec$ representing the 
transverse separation between the emitting regions (in the pulsar's magnetosphere)
corresponding to the leading and trailing edges of the pulsar profile and the
angle $\gamma$ that it makes with the proper motion vector. 
The phase shift of the fringes between the leading and trailing edges of the
profile can be expressed in terms of these quantities as 
\begin{equation}
\deltaph ~=~ \frac {2 \ \pi \; \nut \; \deltaS \; cos(\gamma-\alpha)} 
{\Lp \; \mupm \; cos(\alpha)} ~.
\end{equation}
Here $\alpha$ is the angle between $\mupm$ and the resultant of $\thetaonevec$ 
and $\thetatwovec$, as shown in Figure 4(b).
It is important to note that the above equation is independent of the exact 
location of the screen.  This is because the lever arm effect of $\deltaS$ is 
exactly compensated for by the lever arm effect of the pulsar's motion.

We note that our derivation of equations (6) and (9) ignores the contribution
of the difference in the dispersive time delay along the two paths, as described
by equation (6) of Rickett et al. (1997).  This is because it is easily shown
(as has been done by Rickett et al. 1997 for their data) that the contribution of 
this term to the total frequency dependent delay is negligible.  For example, if 
we assume that the observed value of $\nuf$ is entirely due to differential 
dispersive delay, it would require a fractional DM variation of 
$8.5 \times 10^{-6}$ between the two paths.
Using a typical value of $1 \times 10^{-6}$ for fractional 
DM variations over one year time scales, reported for nearby pulsars by Phillips 
\& Wolszczan (1991), we expect a fractional variation of $5.5 \times 10^{-9}$ over 
the time scale of our event.  Clearly, this is about 3 orders of magnitude smaller 
than required.

Assuming $\Lp$, $\fc$ and $\mupm$ to be known quantities, there are 6 independent 
unknown quantities $-$ $\thetaone$, $\thetatwo$, $\deltaS$, $\alpha_1$, $\alpha_2$ 
and $\gamma$ $-$ and only 3 independent equations $-$ equations (9),(10) and (11).  
Thus, extra input is needed to invert the problem and obtain a solution for $\deltaS$.
Some of this extra information is obtained from the slope of the drift bands produced 
by the two images contributing to the dynamic spectra, by using the relation between 
the drift slopes, the refractive angles and the proper motion.  The two relevant 
equations are 
\begin{equation}
\frac {dt} {df} \biggl |_{1} ~~=~~ -~ \frac {2 \; \thetaone \;cos(\alpha_1)} {\fc \; \mupm}  ~~~{\rm and}~~~ 
\frac {dt} {df} \biggl |_{2} ~~=~~ -~ \frac {2 \; \thetatwo \;cos(\alpha_2)} {\fc \; \mupm}  ~~.
\end{equation}

It is interesting to note that equations (10) and (12) can be used to constrain 
the distances to the pulsar and the scattering structure.  
For this, using equation (12), we substitute for $\thetaone \ cos(\alphaone)$
and $\thetatwo \ cos(\alphatwo)$ in equation (10), to get
\begin {equation}
\frac {\Ls ~ \Lp} {\Lp-\Ls} ~~=~~ 918 ~\pm~ 153~{\rm pc} ~~,
\end{equation}
for the known values of $\fc$, $\mupm$, $\nut$ and the two drift slopes.
The error bar on the above estimate takes into account the errors in the proper
motion of the pulsar and the errors in the estimated drift slopes.  Of these,
the latter are the dominant source.
A range of combinations of $\Ls$ and $\Lp$ values can be found that satisfy this
relationship.
For example, if the scattering structure is assumed to be located midway to the
pulsar ($\Ls \ = \ \Lp/2$), then the distance to the pulsar needs to be 3.4 times
the current estimate of 270 pc.  This is rather unlikely.  On the other hand, if 
the current distance estimate is taken to be correct, it requires the scattering 
structure to be located at $\Ls \ = \ (0.77 ~\pm~ 0.2)~\Lp$.  The error in this 
estimate is dominated by the error in the distance estimate (taken to be 20\%) 
for this pulsar. 
We note that this estimate of the distance to the scatterer is strikingly 
close to the distance to the boundary of the Local Bubble in this direction, as 
inferred by Bhat, Gupta \& Rao (1998) from their detailed study of pulsar 
scintillation in the local ISM.  Their model predicts that the cavity of the 
Local Bubble is surrounded by an ellipsoidal shell of enhanced scattering material.
Along the line of sight to this pulsar, this shell is estimated to contribute
about 70\% of the total scattering.
From their model, the distance to the shell in this direction is expected to 
be in the range 212 to 217 pc ($0.79~\Lp$  to $0.80~\Lp$). 
Thus, it is very likely that the scattering structure responsible for this
fringing event is located in the shell of the Local Bubble.

Using equations (9), (11), (12) and the measurement of the phase shift of the
fringes between the leading and trailing edges of the profile, constraints can 
be placed on the minimum value of $\deltaS$ in the following manner.
For a given choice of one of the unknowns (say, $\thetaone$), unique 
values can be calculated for $\alpha_1$, $\thetatwo$, $\alpha_2$ (and hence 
$\alpha$ and \thetadiff).  Then, assuming the extreme case of $\gamma = \alpha$ a 
lower limit to $\deltaS$ for the given choice of $\thetaone$ can be calculated as 
\begin{equation}
\deltaSmin ~~=~~ \frac {\deltaph \; \Lp \; \mupm \; cos(\alpha)} {2 \; \pi \; \nut} ~~.
\end{equation}
This procedure has been followed for a plausible range of values for $\thetaone$ 
given by $\mid \thetaone \mid \;<\; 20 \; \theta_d $.  A value of $\deltaph ~=~$ 
15\degsp has been used in these calculations. The results are summarized
in Figure 5, where various relevant quantities are plotted against \thetaone.  
From these, it can be seen that the inferred value of 
$\deltaSmin$ ranges from $\approx 1 \times 10^6$ m to $\approx 1 \times 10^5$ m for 
$\thetaone$ ranging from $\approx 1$ mas to $\approx 15$ mas.  The inferred value of 
$\thetatwo$ follows $\thetaone$ closely and the angles $\alpha_1$, $\alpha_2$ 
systematically increase in amplitude towards 90\degsp (but with opposite signs) with 
increasing $\thetaone$.  Consequently, the resultant angle $\alpha$ increases from 
$\approx$ 55\degsp to $\approx$ 90\degsp with increasing $\thetaone$.  This means 
that the smaller values of $\deltaSmin$ are obtained when the orientation of the 
scattering structure is close to being orthogonal to the direction of proper motion 
of the pulsar.


\section{Discussion}

Our observations and analysis of the phase shift of the fringes across the
pulse window clearly reveal that the pulsar magnetosphere has been resolved during 
these observations.  There are two significant aspects of the variation of the fringe 
phase shown in Figure 3.  The first is the inferred separation between the emission
regions corresponding to the leading and trailing edges of the pulsar profile, when
they are radiating towards the observer (i.e., the total size of the emission region). 
The second is the implications of the non-monotonic variation of the fringe phase 
as a function of pulse longitude.

Addressing the first aspect, we first note that the diffractive scattering angle 
($\theta_d$) for this epoch is estimated to be $ \approx 0.72 $ mas.  The standard
Kolmogorov model for the spectrum of electron density fluctuations in the ISM
predicts that the rms value of the refractive scattering angle should be 
\la $\theta_d$ (e.g. Rickett 1990).  Assuming that individual realizations of the 
refractive angles could be $ \approx $ 3$-$5 times larger than the rms values, 
$\thetaone$ and $\thetatwo$ values of $\approx $ 2$-$4 mas are feasible.  From 
Figure 5, this gives the likely value for $\deltaSmin$ to be $\approx 3 \times 10^5$ m.
Smaller values of $\deltaSmin$ require appreciably larger values of $\thetaone$ and 
$\thetatwo$ and are much less likely for the simple Kolmogorov model.  Of course, if 
the spectrum is steeper than Kolmogorov or has a low-frequency enhancement 
corresponding to refractive scales, larger values of $\thetaone$ and $\thetatwo$ would 
be feasible.  Evidence for 
such enhanced refraction has been reported for several different lines of sight 
(e.g. Rickett et al. 1997, Fiedler et al. 1994).  In particular, for this line of 
sight, Bhat, Gupta \& Rao (1999b) find that the slope of the spectrum, estimated from 
diffractive and refractive scintillation measurements, is in excess of 11/3, being
close to 3.78.
In any case, Figure 5 indicates that $\deltaSmin$ appears to reach an asymptotic 
value of $\approx ~8 \times 10^4$ m for large values of $\thetaone$.  
Thus it is unlikely that a value of $\deltaSmin~=~3 \times 10^5$ m would be more 
than a factor of 2 larger than the smallest value that more complicated 
versions of the Kolmogorov model would allow.

From this estimate of $\deltaS$, the altitude at which the observed radiation is
emitted in the pulsar magnetosphere can be calculated.  This requires two assumptions
to be made: the first is that the pulsar's magnetic field has a simple dipole geometry
and the second is that the radiation corresponding to the leading and trailing edges 
of the profile is emitted at the same height and is beamed tangential to field lines 
located symmetrically about the magnetic axis.  This geometry is illustrated in 
Figure 5 of Cordes, Weisberg \& Boriakoff (1983).  Under these conditions, it can be shown that the 
emission altitude is given by
$r_{em} \ \approx \ {3 \ \deltaS} / {\Delta\theta} ~,$
where $\Delta\theta$ is the observed pulse width between the leading and trailing
edges of the profile, expressed in radians.  
The derivation of this result requires the small angle approximation to be valid 
for the angle between the tangent to the field line and the magnetic axis of the pulsar.
Using $\deltaS ~=~ \deltaSmin$ gives a lower limit on $r_{em}$ of $2.6 \times 10^6$ m.  
This is $4.5\%$ of the value of the light cylinder radius for this pulsar.

Our inferred values of $\deltaSmin \approx 3 \times 10^5$ m and $r_{em} \approx 3 \times 10^6$ m 
are the smallest lower limits of all currently available estimates using the technique 
of multiple imaging (Kuz'min 1992, Wolszczan \& Cordes 1987, Wolszczan et al. 1988).
Furthermore, it is the only value that is also consistent with the upper 
limit of $\deltaS < 1 \times 10^{6} $ m obtained for this pulsar from diffractive 
scintillation techniques (Cordes et al. 1983).  We also note that our values are
very similar to those obtained for this pulsar by Smirnova \& Shishov (1989), from 
an analysis of interstellar scintillations at 102 MHz.
Further, our estimates are very similar to the size of the emission region for the 
Vela pulsar ($ \approx $ 500 km) obtained by Gwinn et al. (1997) from VLBI observations 
of the scintillation pattern.  Thus there appears to be some amount of recent evidence 
from scintillation studies that transverse extents of pulsar emission regions may be 
$ \sim \ 10^5 $ m.

Though our inferred values for emission altitude are among the smallest reported values 
obtained from the scintillation technique, they are still somewhat larger than those 
obtained from other techniques for estimating emission altitudes.  For example, 
period-pulse width relation studies (e.g. Rankin 1990, Gil 1991) suggest emission 
altitudes to be $\approx 1 \times 10^5$ m to $1 \times 10^6$ m.  
A similar range is estimated from the technique of Blaskiewicz, Cordes \& Wasserman (1991), 
where measurement of the time lag between the total intensity and polarization position 
angle profiles is used to estimate the emission altitude.  Emission altitudes estimated 
from the measurement of times of arrival for different frequencies using pulsar timing
measurements, place an upper limit of $3\%$ of the light cylinder radius (Cordes 1992; 
Phillips 1992).
More recently, Kijak \& Gil (1998) and Kijak \& Gil (1997) have done a thorough study
of the estimation of the emission altitudes for several pulsars using a refined version
of the period-pulse width relationship.  They have estimated the emission altitudes for 
16 long period pulsars and 6 millisecond pulsars at different frequencies ranging from 
a few 100 MHz to 10 GHz.  From the results for PSR B1133+16 given in Fig. 2 of 
Kijak \& Gil (1998), the emission altitude at 327 MHz can be estimated as 
$7.5 \times 10^5$ m, which is only about 3 times smaller than our estimate.
Thus, the new results presented here have somewhat reduced the discrepancy that 
prevails between the results from the scintillation technique and the other methods 
of estimating pulsar emission altitudes. 

The observed non-monotonic variation of the fringe phase as a function of pulse 
longitude implies, to a first order, that the inferred transverse separations of 
the emission regions do not increase monotonically across the pulsar profile. 
Similar non-monotonic variations of fringe phase have been observed for this pulsar 
(Wolszczan et al. 1988) and for PSR B1237+25 (Wolszczan \& Cordes 1987).  
In the work of Smirnova \& Shishov (1989) and Smirnova, Shishov \& Malofeev (1996), the 
inferred transverse separations as a function of pulse longitude also show a 
non-monotonic variation for several pulsars (including PSR B1133+16).  Thus such 
behaviour does not appear to be an uncommon feature.

There are two models that can explain such results.  The first is the case where
the magnetic field in the emission regions is not strictly dipolar.  The other 
possibility is that the emission altitudes at a given frequency are not constant
across the polar cap.  The shape of the polarization angle curve can be used to 
discriminate between these two possibilities.  If the polarization angle
curve follows the ``S shape'' curve as described by the Radhakrishnan \& Cooke (RC) 
model for a dipole geometry (Radhakrishnan \& Cooke 1969), then the first model 
can be ruled out.
For PSR B1133+16, the observed polarization angle curve does not appear to follow
such a model.  In fact, the shape of the polarization curve is strikingly similar 
to the fringe phase curve.  This can be seen by comparing the fringe phase curve in
Figure 3 with the polarization angle curve at the nearby frequency of 408 MHz 
(e.g. Fig. B.24 of Gould 1994).
Similarly, the fringe phase curve obtained for this pulsar by Wolszczan et al. (1988) 
for the fringing episode at a frequency of 1400 MHz, though very different from our 
curve in Figure 3, is very similar to the position angle curve at 1400 MHz 
(e.g. Fig. 16 of Stinebring et al. 1984 \& Fig. B.24 of Gould 1994).
We note in particular that the longitudes at which the position angle deviates 
by large amounts from a smooth, monotonic curve are also the longitudes where the 
fringe angle curve shows significant deviations.
It is very unlikely that such a similarity between the fringe angle curves at two 
different frequencies and their respective position angle curves is coincidental.

Though the deviation from the RC model curve for this pulsar may, at first sight, 
indicate that the field is not dipolar, 
it is now well understood that the observed deviation is due to the phenomenon of 
orthogonal polarization modes (e.g. Stinebring et al. 1984).  In this picture, the 
position angle at any longitude can take on one of two values at any instant $-$ the 
normal position angle value and a value 90\degsp different from this.  The longitudes 
at which the secondary, orthogonal mode occurs more frequently will show a mean 
polarization angle that is about 90\degsp different from the overall position angle 
curve produced by the primary mode.  Such an effect is clearly seen in part of the 
first component of the profile for this pulsar, in the 1400 MHz data 
(Stinebring et al. 1984).  
For longitudes where the dominance of the primary mode over the secondary mode is 
not well established, the deviations of position angle are not as dramatic. This 
could give rise to the smoother (but still deviating from the RC model) polarization 
angle curve seen at 400 MHz for this pulsar.
On plotting the mode separated position angle curves, each is found to follow 
the RC model, while maintaining a separation of 90\degsp.  
It is not clear whether orthogonal polarization states are intrinsic to the pulsar 
emission mechanism or are an artifact of propagation through the magnetospheric plasma.

Since the fringe angle curve shows significant deviations from a monotonic curve 
at the same longitudes at which the secondary polarization mode is dominant, it 
suggests that the emission regions for the secondary mode are significantly offset
in location from the monotonically increasing separation of the emission
regions representing the primary mode.  However, since both modes at a given longitude
are expected to originate at the same field line, this offset in separation is 
unlikely to be along the polar cap (at a constant altitude).  Rather, this offset 
is more likely produced by a difference in emission altitudes for the two modes, 
while following the same field line.  
This suggests that the orthogonal polarization modes are likely to be a propagation 
effect rather than an intrinsic property of the emission mechanism.  

Our results and inferences are in agreement with models where refraction in the 
pulsar magnetosphere is used to explain the phenomenon of orthogonal polarization 
modes.  For example, Barnard \& Arons (1986) have shown that the relativistically 
outflowing plasma in the open field line region of the pulsar magnetosphere can 
act as a ``polarizing beam splitter''. 
In their model, X-mode and O-mode waves, having intrinsically orthogonal polarization
states, are produced that get separated by large angles due to the fact that
the X-mode wave travels in a straight line from the point of emission, while the
O-mode wave is ducted along the magnetic field line, to a point higher in the 
magnetosphere from where it escapes and follows a straight line trajectory. 
Gallant (1996) has shown how a combination of such X-mode waves and ducted O-mode 
waves can produce the non-monotonic variation of the lateral displacement 
of the emission regions across the pulse profile, as inferred in the scintillation
studies.  Such a model would naturally predict a similarity between
the phase angle curves of the fringing events and the polarization angle curve.

Further data of multiple imaging events (preferably multiple epochs for a pulsar) 
with careful measurements of drift slopes in the dyanmic spectra should provide 
better constraints on the pulsar emission geometry.  Such observations might also 
help to infer properties of the magnetosphere, such as the plasma density, and 
help in studying propagation effects in pulsar magnetospheres.


\section{Conclusions}

We have reported the observations and analysis of a multiple imaging event for
PSR B1133+16 at a frequency of 327 MHz.  The dynamic spectra show clear 
periodic intensity modulations as a function of time and frequency. In addition, 
dual drift slopes are clearly seen in the dynamic spectra. The fringes are seen
in each of the 10 bins across the pulse profile.  The phase of the fringes shows
a change of $\approx$ 15\degsp between the leading and trailing edges of the 
pulsar profile, implying that the pulsar magnetosphere has been resolved during 
these observations. However, the observed fringe phase variation with pulse 
longitude is non-monotonic.
We have modelled the  observed fringes as being produced by multipath propagation 
due to refraction from a discrete scattering object in the ISM.  
Expressions have been derived for the dual drift slopes, the fringe 
periodicities in time and frequency and the transverse separation of emission 
regions as a function of the measured fringe phase shift.
From the measurements of the fringe period in time and the values of the 
dual drift slopes, we conclude that the refracting structure is located at 
a distance that matches very well with the expected location, in this direction,
of the enhanced scattering shell of the Local Bubble.  Thus it is very likely that
the enhanced scattering required to produce this multiple imaging is provided by
the shell of the Local Bubble. 
Using our model, we are able to constrain the minimum size of the emission region, 
$\deltaSmin$, to be $ \approx \ 3 \times 10^5 $ m. 
For a dipole geometry, this implies an emission altitude in the magnetosphere of 
$ \approx \ 2.6 \times 10^6$ m, which is $4.5\%$ of the light cylinder radius for
this pulsar.
These estimates of $\deltaSmin$ and emission altitude provide the smallest lower 
limits of all currently available estimates using the technique of multiple imaging.  
The estimated emission altitude is however larger than that obtained using other 
techniques such as multifrequency timing, polarization and period-pulse width 
relation studies.
Refraction in the pulsar magnetosphere is the likely explanation for this disagreement.
The non-monotonic variation of the fringe phase with pulse longitude for this pulsar
matches very closely with the variation of the polarization angle curve with pulse 
longitude.  This is found to be true at two different radio frequencies. 
This is interpreted as implying that the orthogonal polarization modes for this 
pulsar are emitted at different altitudes in the pulsar magnetosphere.
Such a picture is in agreement with models where refraction plays an important role 
in the pulsar magnetosphere.

 
{}


\clearpage

\section{Figure captions}
 
\begin{description}
 
\item [Figure 1]
Dynamic scintillation spectra of PSR B1133+16 at 3 epochs. 
The intensity variations are represented in a linear grey scale, 
with darker regions corresponding to higher intensity values.
The date of the observation is given at top right of each panel 
and the mean flux density at the top left of each panel. 

\item [Figure 2]
Dynamic spectrum for the second epoch (30 April 1994), after passing through
high pass filter (top panel) and low pass filter (middle panel), as described
in the text.  The bottom panel shows the auto-correlation functions for the
low pass filtered data, as described in the text.  The solid and dashed lines
represent the estimated drift slopes.

\item [Figure 3]
Variation of amplitude and phase of the periodic signal detected in the
dynamic spectrum of the second epoch, as a function of longitude bin number.

\item [Figure 4]
Schematics illustrating the geometry of the scattering event for 
(a) plane containing the pulsar (P), observer (O) and scattering structure, 
and (b) plane transverse to the line of sight of the pulsar.

\item [Figure 5]
Variation of the different quantities as a function of refractive scattering angle, $\theta_{r1}$.  
In the middle panel, the solid line is for $\theta_{r2}$ and the dashed line is for $\thetadiff$. 

\end{description}
 


\begin{thebibliography}{}

\bibitem{} Barnard, J. J. \& Arons, J. 1986, ApJ, 302, 138

\bibitem{} Bhat, N. D. R., Gupta, Y. \& Rao, A. P. 1998, ApJ, 500, 262

\bibitem{} Bhat, N. D. R., Rao, A. P. \& Gupta, Y. 1999a, ApJ, In Press

\bibitem{} Bhat, N. D. R., Gupta, Y. \& Rao, A. P. 1999b, ApJ, In Press

\bibitem{} Blaskiewicz, M., Cordes, J. M. \& Wasserman, I. 1991, ApJ, 370, 643

\bibitem{} Cordes, J. M., Weisberg, J. \& Boriakoff, V. 1983, ApJ, 268, 370

\bibitem{} Cordes, J. M. \& Wolszczan, A. 1986, ApJ, 307, L27

\bibitem{} Cordes, J. M. \& Wolszczan, A. 1988, 
in AIP Conf. Proc. No. 174 - Radiowave Scattering in the Interstellar Medium,
ed. Cordes, J. M., Rickett, B. J. \& Backer, D. C.
(New York: AIP), 212

\bibitem{} Cordes, J. M. 1992, 
in Proceedings of the IAU Colloquium No. 128
`` The Magnetospheric Structure and Emission Mechanisms of Radio Pulsars'',
eds. T. H. Hankins, J. M. Rankin \& J. A. Gil,
(Zielona Gora: Pedagogical University Press), 253

\bibitem{} Fiedler, R., Dennison, B., Johnston, K. J., Waltman, E. B. 
\& Simon, R. S. 1994, ApJ, 430, 581

\bibitem{} Gallant, Y. A.  1996, 
in Proceedings of the IAU Colloquium No. 160 
``Pulsars: Problems and Progress'', 
eds. S. Johnston, M. A. Walker, \& M. Bailes, 
(San Francisco: Astron. Soc. of the Pacific), 431.

\bibitem{} Gil, J. A. 1991, A\&A, 243, 219

\bibitem{} Gould, D. M. 1994, Ph.D. thesis, University of Manchester

\bibitem{} Gupta, Y., Rickett, B. J. \& Lyne, A. G. 1994, MNRAS, 269, 1035

\bibitem{} Gupta, Y. 1995, ApJ, 451, 717

\bibitem{} Gwinn, C. R., et al. 1997, ApJ, 483, L53

\bibitem{} Hewish, A. 1980, MNRAS, 192, 799

\bibitem{} Kijak, J. \& Gil, J. 1997, MNRAS, 288, 631

\bibitem{} Kijak, J. \& Gil, J. 1998, MNRAS, 299, 855

\bibitem{} Kuz$^{,}$min, O. A. 1992,
in Proceedings of the IAU Colloquium No. 128
`` The Magnetospheric Structure and Emission Mechanisms of Radio Pulsars'',
eds. T. H. Hankins, J. M. Rankin \& J. A. Gil,
(Zielona Gora: Pedagogical University Press), 287

\bibitem{} Phillips, J. A. \& Wolszczan, A. 1991, ApJ, 382, L27

\bibitem{} Phillips, J. A. 1992, ApJ, 385, 282

\bibitem{} Radhakrishnan, V. \& Cooke, D. J. 1969, Astrophy. Lett., 3, 225

\bibitem{} Rankin, J. M. 1990, ApJ, 352, 247

\bibitem{} Rickett, B. J. 1990, ARA\&A, 28, 561

\bibitem{} Rickett, B. J. 1996, 
in Proceedings of the IAU Colloquium No. 160
``Pulsars: Problems and Progress'',
eds. S. Johnston, M. A. Walker, \& M. Bailes,
(San Francisco: Astron. Soc. of the Pacific), 439

\bibitem{} Rickett, B. J., Lyne, A. G. \& Gupta, Y. 1997, MNRAS, 287, 739

\bibitem{} Smirnova, T. V. \& Shishov, V. I. 1989, Pis'ma AZh, 15, 443; English transl.
Soviet Astron., 15, 191

\bibitem{} Smirnova, T. V., Shishov, V. I. \& Malofeev, V. M. 1996, ApJ, 462, 289

\bibitem{} Stinebring, D. R., Cordes, J. M., Rankin, J. M., Weisberg, J. M. \& 
Boriakoff, V. 1984, ApJSS, 55, 247

\bibitem{} Wolszczan, A. \& Cordes, J. M. 1987, ApJ, 320, L35

\bibitem{} Wolszczan, A., Bartlett, J. E. \& Cordes, J. M. 1988, 
in AIP Conf. Proc. No. 174 - Radiowave Scattering in the Interstellar Medium,
ed. Cordes, J. M., Rickett, B. J. \& Backer, D. C.
(New York: AIP), 145

\end{thebibliography}
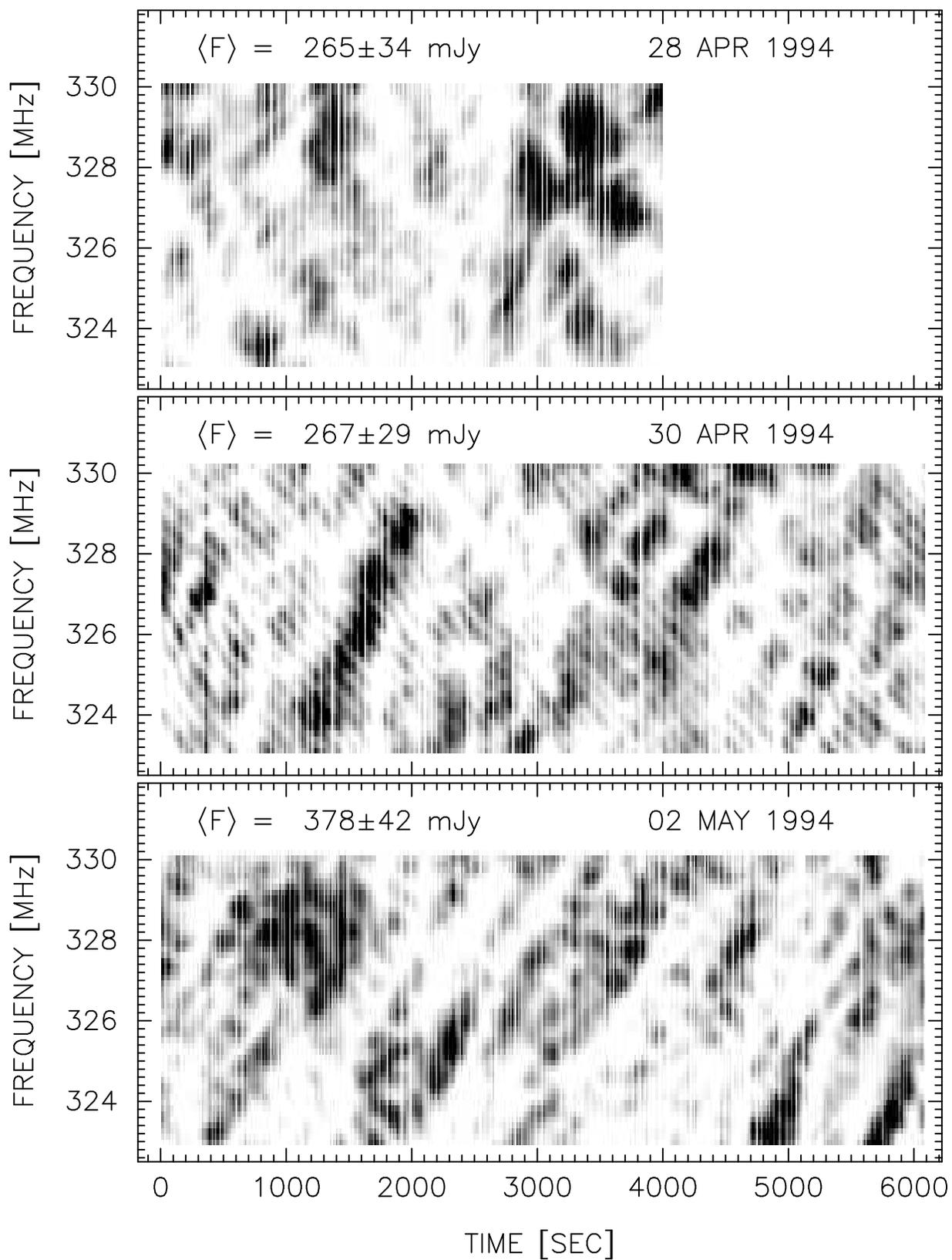
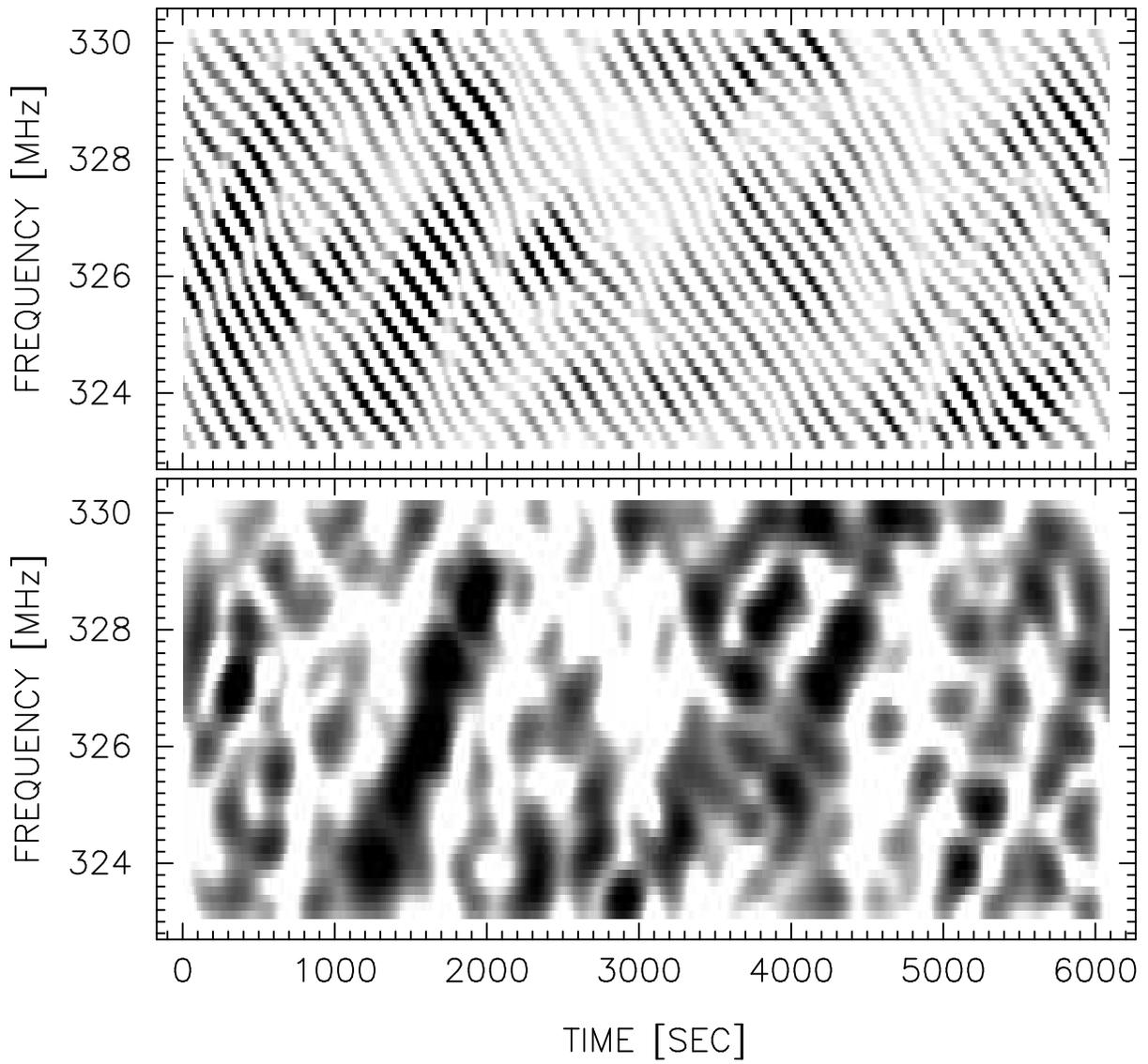
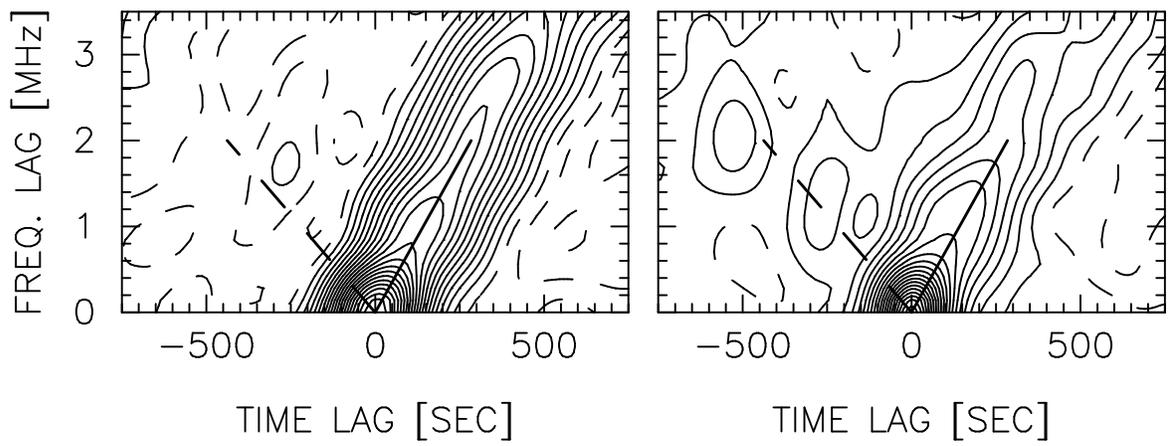
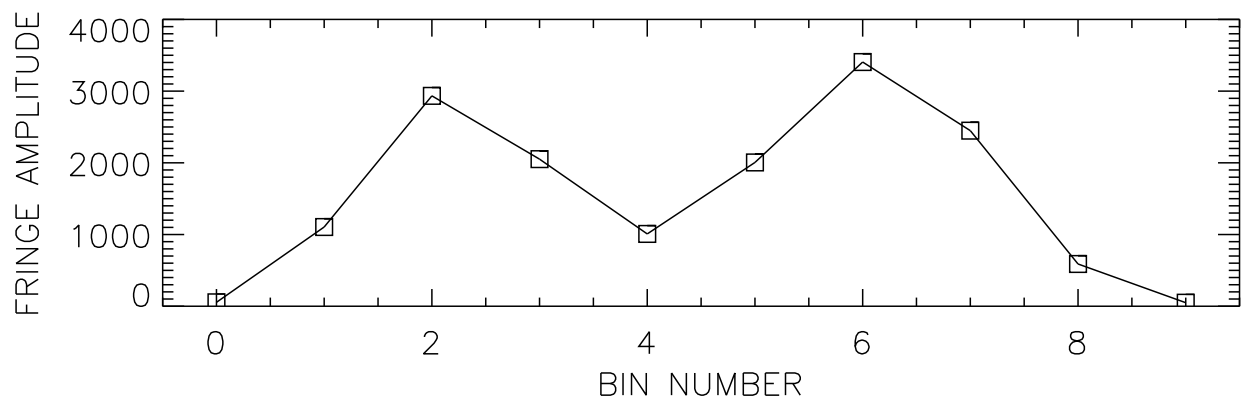
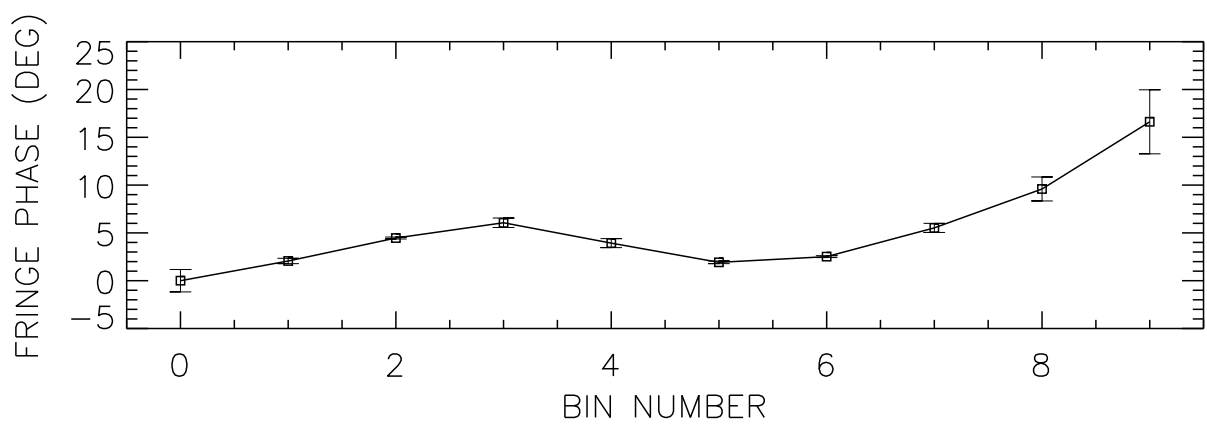
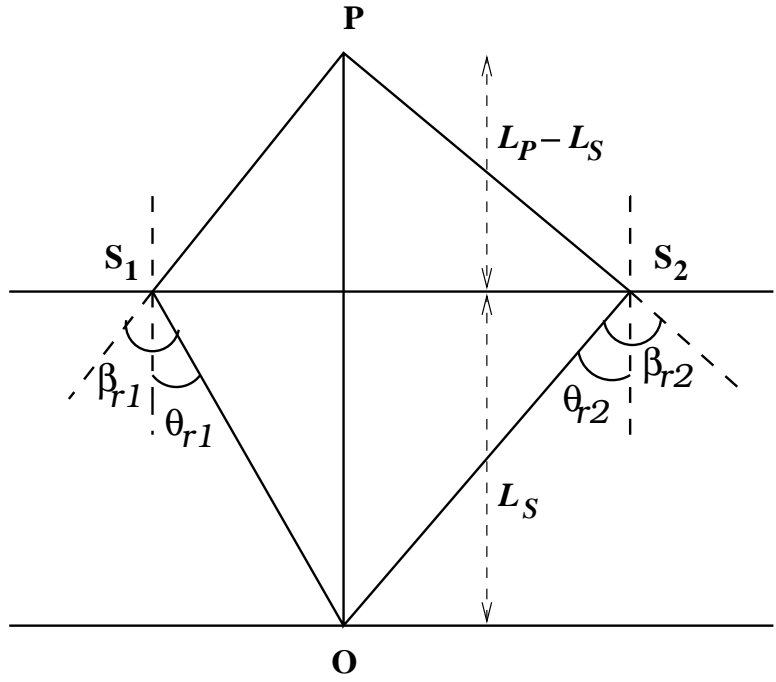
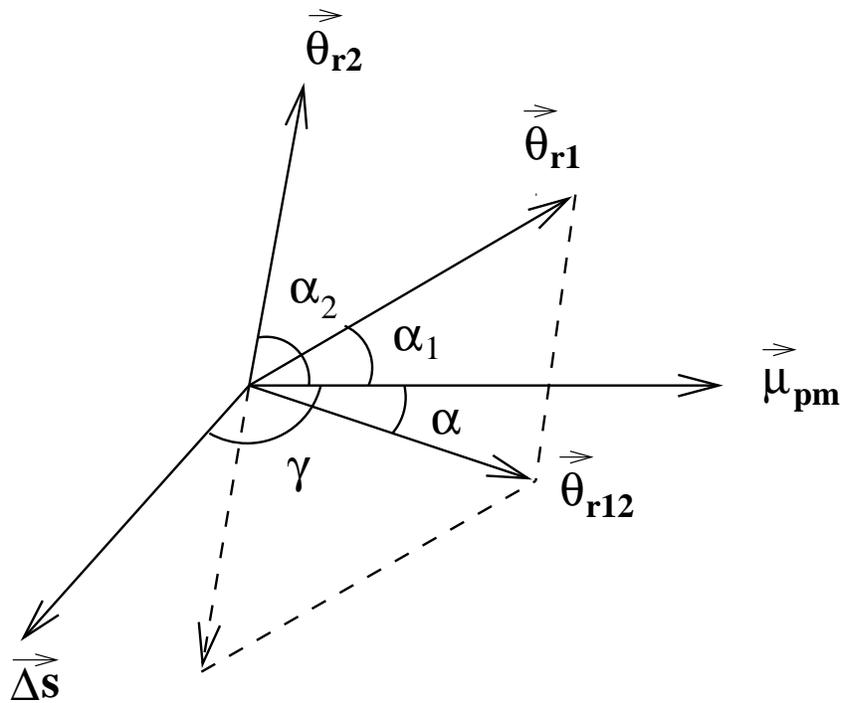
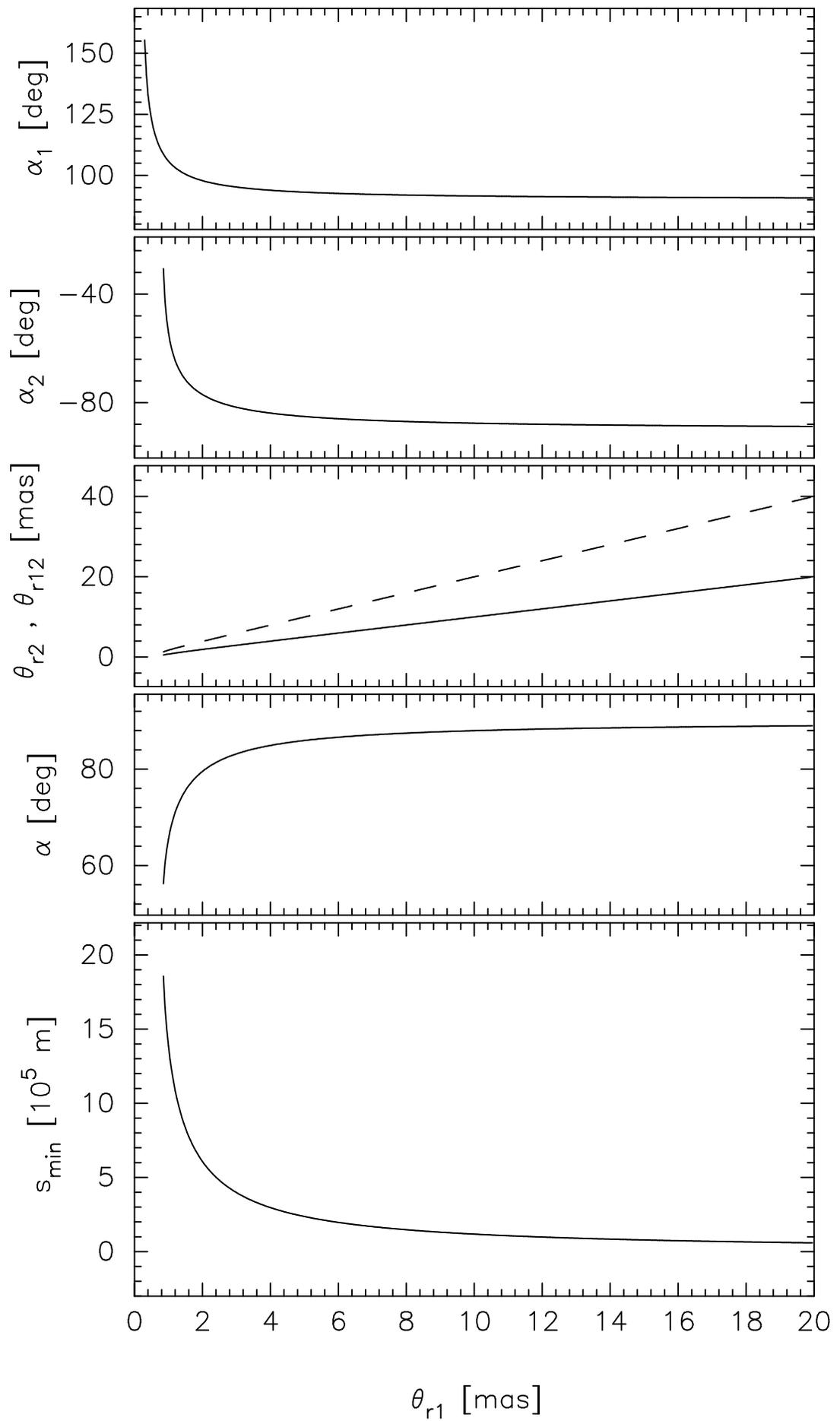
\end{document}